\documentclass[10pt, conference, compsocconf]{IEEEtran}

\usepackage{times}
\usepackage{amsmath}
\usepackage{stmaryrd}
\usepackage{amssymb}
\usepackage{graphicx}
\usepackage[T1]{fontenc}

\usepackage[ruled]{algorithm2e}

\def\>{\ensuremath{\rangle}}
\def\<{\ensuremath{\langle}}

\newtheorem{thm}{Theorem}[section]

\newtheorem{lem}{Lemma}[section]
\newtheorem{defn}{Definition}[section]

\newtheorem{exam}{Example}[section]

\newtheorem{rem}{Remark}[section]

\begin{document}
\title{Symbolic Verification of Quantum Circuits}

\author{\IEEEauthorblockN{Mingsheng Ying$^{1,2,3}$ and Zhengfeng Ji$^1$}
\IEEEauthorblockA{1. Centre for Quantum Software and Information, 
University of Technology Sydney,
Australia\\ 2. State Key Laboratory of Computer Science, Institute of Software, Chinese Academy of Sciences, China\\ 3. Department  of Computer Science and Technology, Tsinghua University, China\\
Emails: Mingsheng.Ying@uts.edu.au; Zhengfeng.Ji@uts.edu.au}
}
\maketitle

\begin{abstract}
This short note proposes a symbolic approach for representing and reasoning about quantum circuits using complex, vector or matrix-valued Boolean expressions. A major benefit of this approach is that it allows us to directly borrow the existing techniques and tools for verification of classical logic circuits in reasoning about quantum circuits. 
\end{abstract}


\IEEEpeerreviewmaketitle

\section{Introduction}\label{Intro}

Several formal methods for reasoning about quantum circuits have been proposed  in the last ten years, with potential applications in verification of quantum hardware \cite{Pal18} and quantum compilers \cite{Rand}.  
For example, a few variants of BDDs, including QuIDD \cite{Via07}, QMDD \cite{Sei12} and TDD \cite{Hong} have been introduced for equivalence checking of quantum circuits as well as other verification issues. Reversible miter was introduced in \cite{Markov} so that various simplification techniques of quantum circuits can be used in equivalence checking.
A small set of laws for some basic operations on vectors and matrices was identified in \cite{Shi} in order to avoid as much as possible explicit computation of matrices through symbolic manipulation.

In this short note, we introduce a new approach for formally reasoning about quantum circuits. It is mainly motivated by the following observation:\begin{itemize}
\item The designer of a classical circuit usually has in mind a certain logical structure of the variables and gates used in the circuit.
There is no difference in designing a quantum circuit. 
But quantum states and quantum gates are usually explicitly represented by complex vectors and matrices, respectively, in which the logical structure is totally lost.  
\end{itemize}

This approach is built upon a symbolic representation of quantum states and quantum gates using so-called matrix-valued Boolean expressions (including complex number-valued and vector-valued Boolean expressions). In particular, many quantum gates are naturally defined by a couple of case statements. These gates can be more naturally represented by Boolean expressions than as matrices (for instance, see Example \ref{Deutsch}).     
Then verification of quantum circuits can be done by a flexible combination of the following two layers of reasoning:
\begin{itemize}\item Classical logical reasoning at the bottom; thus all of the existing techniques for reasoning about \textit{classical} circuits can be used;
\item Manipulation and operation of complex numbers or matrices at the top. Here, special properties (e.g. unitarity) of the involved matrices (e.g. Hadamard gate, Pauli gates, controlled gates) can be utilised.     
\end{itemize}
It is worth noting that this idea actually coincides with human's reasoning about the correctness of quantum circuits. 

A benefit of our approach is that it provides us with a tool that can retrieve and leverage the logical structure underlying a quantum circuit in reasoning about it. 
A large variety of effective techniques and automatic tools have been developed for verification of classical logic circuits, with successful industrial applications.  
Perhaps a more important benefit of our approach is that these techniques and tools can be directly \textit{reused} in verification of quantum circuits. 

\section{Motivating Examples}

To illustrate our basic idea, let us consider the following two simple examples. The reader can understand the notations used in these examples according to her/his intuition, and their precise definitions will be given in the later sections.  

{\vskip 3pt}

\begin{exam}\label{ex-in1}Our first example shows that the Hadamard gate can generate a superposition. Let $q$ be a qubit variable. Then basis states $|0\rangle$ and $|1\rangle$ of $q$ can be expressed as Boolean literals $\overline{q}, q$, respectively, and a general pure state can be written as $$|\psi\rangle=\alpha|0\rangle+\beta|1\rangle=\alpha \overline{q}+\beta q.$$ To describe a quantum gate on $q$, 
we introduce $q^\prime$ as a copy of $q$, and $q, q^\prime$ can be considered as the input qubit and the output qubit of the gate, respectively. Then a general gate on $q$  
can be written as $$U=\left(\begin{array}{cc}u_{00} & u_{01}\\ u_{10} & u_{11}\end{array}\right)=u_{00}\overline{q} \overline{q}^\prime+u_{01}\overline{q}q^\prime+u_{10}q\overline{q}^\prime+u_{11}qq^\prime.$$
An advantage of such an expression is that it can often be simplified using classical Boolean reasoning. For example, the Hadamard gate:\begin{align}
\label{h-matrix}H&=\frac{1}{\sqrt{2}}\left(\begin{array}{cc}1 & 1 \\ 1 & -1\end{array}\right)\\ &=\frac{1}{\sqrt{2}}(\overline{q}\overline{q}^\prime+\overline{q} q^\prime+q\overline{q}^\prime-qq^\prime)\\
\label{h-intui}&=\frac{1}{\sqrt{2}}[(\overline{q}+\overline{q}^\prime)-qq^\prime].
\end{align} It is particularly interesting that logical expression (\ref{h-intui}) is closer to our intuition about the Hadamard gate than matrix (\ref{h-matrix}): $|1\rangle$ is transformed to $|1\rangle$ with amplitude $-\frac{1}{\sqrt{2}}$, and the amplitude is always $\frac{1}{\sqrt{2}}$ in other cases. This advantage can be seen even more clearly when we compute the output of the Hadamard gate with basis $|0\rangle$ as its input; that is, a superposition $$|+\rangle=\frac{1}{\sqrt{2}}(|0\rangle+|1\rangle)$$ of basis states $|0\rangle$ and $|1\rangle$ is generated by the Hadamard gate from $|0\rangle$:\begin{align}
H|0\rangle&=\exists q:\frac{1}{\sqrt{2}}[(\overline{q}+\overline{q}^\prime)-qq^\prime]\overline{q}\\
\label{h-mid}&=\exists q:\frac{1}{\sqrt{2}}[(\overline{q}+\overline{q}^\prime)\overline{q}-qq^\prime \overline{q}]\\
\label{h-simp}&=\exists q:\frac{1}{\sqrt{2}}\overline{q}\\ &= \frac{1}{\sqrt{2}}= \frac{1}{\sqrt{2}}(\overline{q}^\prime+q^\prime)=\frac{1}{\sqrt{2}}(|0\rangle+|1\rangle).
\end{align} Note that (\ref{h-simp}) is obtained from (\ref{h-mid}) by Boolean reasoning that $(\overline{q}+\overline{q}^\prime)\overline{q}=\overline{q}$ and $qq^\prime \overline{q}=(q\overline{q})q^\prime =0$. 
\end{exam}

{\vskip 3pt}

\begin{exam}\label{ex-in2} Our second example shows that the CNOT (controlled-NOT) gate can create an entanglement or EPR (Einstein-Podolsky-Rosen) pair. Let $q_1,q_2$ be two qubit variables. We introduce their output copies $q_1^\prime, q_2^\prime$. Then the CNOT gate with $q_1$ as the control qubit and $q_2$ as the target qubit can be written as
\begin{align}\label{cnot-matrix}{\rm CNOT}&=\left(\begin{array}{cccc}1&0&0&0\\ 0&1&0&0\\0&0&0&1\\0&0&1&0\end{array}\right)\\ &=\overline{q}_1\overline{q}_2\overline{q}^\prime_1\overline{q}^\prime_2+\overline{q}_1q_2\overline{q}^\prime_1q_2^\prime+q_1\overline{q}_2q_1^\prime q_2^\prime+q_1q_2q_1^\prime\overline{q}_2^\prime\\ \label{cnot-logic}&=
\overline{q}_1\overline{q}_1^\prime(q_2\leftrightarrow q_2^\prime)+q_1q_1^\prime(q_2\leftrightarrow\overline{q}_2^\prime).
\end{align}
Again, logical expression (\ref{cnot-logic}) is closer to our intuition about CNOT than matrix (\ref{cnot-matrix}): the state of control qubit $q_1$ is not changed, and if $q_1$ is off ($0$), then the state of target qubit $q_2$ is unchanged, but if $q_1$ is on ($1$), then $q_2$ is flipped. Now we input a separate state $$|+\rangle|0\rangle=\frac{1}{\sqrt{2}}(|0\rangle+|1\rangle)|0\rangle=\frac{1}{\sqrt{2}}(\overline{q}_1\overline{q}_2+q_1\overline{q}_2)$$ of $q_1,q_2$ to CNOT, and the output is computed as follows: 
\begin{align*}&{\rm CNOT}|+\rangle|0\rangle =\exists q_1,q_2: \left[\overline{q}_1\overline{q}_1^\prime(q_2\leftrightarrow q_2^\prime)+q_1q_1^\prime(q_2\leftrightarrow\overline{q}_2^\prime)\right]\\ &\qquad\qquad\qquad\qquad\qquad\qquad\qquad\cdot \frac{1}{\sqrt{2}}(\overline{q}_1\overline{q}_2+q_1\overline{q}_2)\\
&=\exists q_1,q_2: \frac{1}{\sqrt{2}}\left[\overline{q}_1\overline{q}_1^\prime(q_2\leftrightarrow q_2^\prime)\overline{q}_2+q_1q_1^\prime(q_2\leftrightarrow \overline{q}_2^\prime)\overline{q}_2\right]\\
&=\frac{1}{\sqrt{2}}(\overline{q}_1^\prime\overline{q}_2^\prime+q_1^\prime q_2^\prime)=\frac{1}{\sqrt{2}}(q_1^\prime\leftrightarrow q_2^\prime)
\\ &=\frac{1}{\sqrt{2}}(|00\rangle+|11\rangle)=|{\rm EPR}\rangle.
\end{align*}
Note that Boolean reasoning is used in the above transformation; for example, $\exists q_1,q_2:q_1q_1^\prime(q_2\leftrightarrow\overline{q}_2^\prime)\overline{q}_2 =\exists q_1: q_1q_1^\prime q_2^\prime=q_1^\prime q_2^\prime.$
\end{exam}

\section{Matrix-Valued Boolean Functions}

The two simple examples in the previous section clearly indicate that classical Boolean logic is really helpful in reasoning about quantum circuits. Now we start to define a formal framework in which the power of Boolean reasoning can be further exploited in verification, simulation and optimisation of quantum circuits.   

\subsection{Boolean Functions}

For convenience of the reader, let us first recall some basic notions about classical Boolean logic. We use $0,1$ to denote the truth values of false, true, respectively. The connectives of not, and, or, exclusive-or, implication and bi-implication are denoted by $^-,\cdot, +,\oplus, \rightarrow, \leftrightarrow$, respectively. We use $x_1,x_2,...$ to denote Boolean variables. A Boolean function with inputs $x_1,...,x_n$ is a mapping: $$F=F(x_1,...,x_n):\{0,1\}^n\rightarrow\{0,1\},$$ and a Boolean expression $f$ over variables $x_1,...,x_n$ is an expression constructed from $x_1,...,x_n$ using the connectives. A Boolean expression determines a Boolean function $F(f)$. For simplicity, we often abuse the notation and use $f$ to denote $F(f)$. The following definition will be frequently used in this note. 

{\vskip 3pt}

\begin{defn}[Cofactor, Quantifications]\label{def-boole-cofactor} Let $F$ be a Boolean function with inputs $x_1,...,x_n$, $1\leq i\leq n$ and $b\in\{0,1\}$. Then:\begin{enumerate}\item The cofactor of $F$ with $x_i=b$ is 
the Boolean function with input $x_1,...,x_{i-1},x_{i+1},...,x_n$:
\begin{align*}&F_{x_i=b}(b_1,...,b_{i-1},b_{i+1},...,b_n)\\ &\qquad\qquad\qquad =F(b_1,...,b_{i-1},b,b_{i+1},...,b_n)\end{align*}
for any $(b_1,...,b_{i-1},b_{i+1},...,b_n)\in\{0,1\}^{n-1}.$

\item The existential and universal quantifications of $F$ over $x_i$ are the Boolean functions with inputs $x_1,...,x_{i-1},x_{i+1},...,x_n$ defined by 
\begin{equation*}\label{quantification}\begin{split}
\exists x_i:F &=F_{x_i=0}+F_{x_i=1},\\
\forall x_i:F &=F_{x_i=0}\cdot F_{x_i=1}.
\end{split}\end{equation*}
\end{enumerate}
\end{defn}

The notion of pseudo-Boolean function has been introduced as a mapping from Boolean values to integers for verification of arithmetic functions (see e.g. \cite{Bryant95}).
There are many different representations of Boolean functions and pseudo-Boolean functions, e.g. truth table, conjunctive and disjunctive normal forms. Among them, BDD (Binary Decision Diagram) plays a dominant role in today's verification tools for classical logic circuits. In particular, various operations of Boolean functions, including cofactor and quantifications can be efficiently implemented in ROBDD (Reduced Ordered Binary Decision Diagram), a canonical form of BDD (see for example \cite{Molit}, Chapter 3).  
 
\subsection{Matrix-Valued Boolean Functions}

Now we can define matrix-valued Boolean functions as a generalisation of Boolean functions and pseudo-Boolean functions. We write $\mathbb{C}$ for the field of complex numbers. For integers $m,k\geq 1$, let $\mathbb{C}_{m\times k}$ stand for the set of $m\times k$ complex matrices. It is convenient to view a complex number as a $1\times 1$ matrix, and  
define the product of Boolean value $b\in\{0,1\}$ and a matrix $A$ as follows: \begin{equation}\label{boole-ring}b\cdot A=A\cdot b=\begin{cases}0\ ({\rm zero\ matrix}) &{\rm if}\ b=0,\\ A &{\rm if}\ b=1.
\end{cases}
\end{equation}

{\vskip 3pt}

\begin{defn}[Matrix-Valued Boolean Function] Let $x_1,...,x_n$ be a sequence of Boolean variables. Then an $m\times k$ matrix-valued Boolean function with inputs $x_1,...,x_n$ is a mapping from the state space $\{0,1\}^n$ of $x_1,...,x_n$ to $m\times k$ matrices:
$$F=F(x_1,...,x_n):\{0,1\}^n\rightarrow \mathbb{C}_{m\times k}.$$ In particular, it is called a (column) vector-valued Boolean function when $k=1$, and a complex-valued Boolean function when $m=k=1$. 
\end{defn}

Definition \ref{def-boole-cofactor} can be straightforwardly generalised to matrix-valued Boolean functions.

{\vskip 3pt}

\begin{defn}[Cofactor, Existential Quantification]\label{Boole-cofactor} Let $F$ be a matrix-valued Boolean function with inputs $x_1,...,x_n$, $1\leq i\leq n$ and $b\in\{0,1\}$. Then:\begin{enumerate}
\item The cofactor of $F$ with $x_i=b$ is the matrix-valued Boolean function with inputs $x_1,...,x_{i-1},x_{i+1},...,x_n$:
\begin{align*}
&F_{x_i=b}(b_1,...,b_{i-1},b_{i+1},...,b_n)\\ &\qquad\qquad\qquad =F(b_1,...,b_{i-1},b,b_{i+1},...,b_n)\end{align*} for any $(b_1,...,b_{i-1},b_{i+1},...,b_n)\in\{0,1\}^{n-1}.$

\item The existential quantification of $F$ over $x_i$ is the matrix-valued Boolean function with inputs $x_1,...,x_{i-1},x_{i+1},...,x_n$ defined by 
\begin{equation}\label{quantification}\begin{split}
\exists x_i:F &=F_{x_i=0}+F_{x_i=1}.
\end{split}\end{equation}
Note that $+$ in the right-hand side of (\ref{quantification}) is the sum of matrices (or more precisely, matrix-valued functions). \end{enumerate}
\end{defn}

{\vskip 3pt}

\begin{rem}The notion of universal quantification can be defined only for $m\times m$ matrix-valued Boolean functions but not for $m\times k$ ones with $m\neq k$. It will not be needed in this note. 
\end{rem}

Using the convention (\ref{boole-ring}), Boolean-Shannon expansion can be generalised to matrix-valued Boolean functions.   

{\vskip 3pt}

\begin{thm}[Boole-Shannon Expansion] For any matrix-valued Boolean function $F$ with inputs $x_1,...,x_n$ and $1\leq i\leq n$, we have:
 $$F=\overline{x}_i\cdot F_{x_i=0}+x_i\cdot F_{x_i=1}.$$
\end{thm}

\section{Matrix-Valued Boolean Expressions}

{\vskip 3pt}

Every Boolean function can be expressed as a (but not unique) Boolean expression. Similarly, we can introduce matrix-valued Boolean expressions for representing matrix-valued Boolean functions.  

{\vskip 3pt}

\begin{defn}[Matrix-valued Boolean expression] Let $x_1,...,x_n$ be Boolean variables. Then an $m\times k$ matrix-valued Boolean expression over $x_1,...,x_n$ is a formula of the form:
\begin{equation}\label{expression}f=\sum_{i=1}^mA_if_i=A_1f_1+\cdots+A_lf_l\end{equation}
where $l\geq 1$, $A_1,...,A_l\in\mathbb{C}_{m\times k}$ and $f_1,...,f_l$ are (classical) Boolean functions over $x_1,...,x_n$.
 \end{defn}
 
 Several useful operations can be defined for matrix-valued Boolean expressions: 
 
 {\vskip 3pt}

\begin{defn}Let $f=\sum_iA_if_i$ be an $n\times m$ matrix-valued Boolean expression and $g=\sum_jB_jg_j$ an $m\times k$ matrix-valued Boolean expression over the same variables $x_1,...,x_n$. Then their product is defined as the $n\times k$ matrix-valued Boolean expressions over $x_1,...,x_n$: $$f\cdot g=\sum_{i,j}(A_iB_j)(f_i\cdot g_j).$$
\end{defn}

{\vskip 3pt}

\begin{defn} Let $f=\sum_iA_if_i$ be a matrix-valued Boolean expression over $x_1,...,x_n$, 
$1\leq k\leq n$ and $b\in\{0,1\}$. Then the cofactor of $f$ with $x_k=b$ and existential quantification of $f$ over $x_k$ are the matrix-valued Boolean expressions over $x_1,...,x_{k-1},x_{k+1}, ..., x_n$: 
\begin{align*}f_{x_k=b}&=\sum_i A_i(f_i)_{x_k=b},\\ \exists x_k:f&=\sum_iA_i(\exists x_k:f_i),\end{align*}
where for each $i$, $(f_i)_{x_k=b}$ and $\exists x_k:f_i$ are defined according to Definition \ref{Boole-cofactor}.\end{defn}

We now see how a matrix-valued Boolean expression can be used to describe a matrix-valued Boolean function.   

{\vskip 3pt}

\begin{defn}A matrix-valued Boolean expression $f=\sum_{i=1}^mA_if_i$ defines a matrix-valued Boolean function $F=F(f)$ by 
$$F(b_1,...,b_n)=\sum\left\{|A_i: f_i(b_1,...,b_n)=1\right|\}$$
for any $(b_1,...,b_n)\in\{0,1\}^n$, where $\{|\cdot|\}$ denotes a multi-set. In particular, if for all $i$, $f_i(b_1,...,b_n)=0$, then $F(b_1,...,b_n)=0$. \end{defn}

The following theorem presents a way of computing the matrix-valued Boolean function of a matrix-valued Boolean expression, and shows that various operations of matrix-valued Boolean expressions are consistent with the corresponding operations of matrix-valued Boolean functions. 

{\vskip 3pt}

\begin{thm}\label{thm-function}\begin{enumerate}\item For any matrix-valued Boolean expression $f=\sum_{i} A_if_i$,
$$F(f)=\sum_{i=1}^m A_i F(f_i)$$ where for each $i$, $F(f_i)$ is the (classical) Boolean function defined by Boolean expression $f_i$, and $A_iF(f_i)$ is defined according to (\ref{boole-ring}).
\item For any matrix-valued Boolean expressions $f,g$, we have: 
\begin{align*}F(f\cdot g)&=F(f)\cdot F(g),\\
F(f_{x_k=b})&=F(f)_{x_k=b},\\ F(\exists x_k:f)&=\exists x_k:F(f).\end{align*}
\end{enumerate}
\end{thm}


The equivalence of two matrix-valued Boolean expressions can be defined in terms of the matrix-valued Boolean functions determined by them.  

{\vskip 3pt}

\begin{defn}Two matrix-valued Boolean expressions $f$ and $g$ are equivalent, written $f\equiv g$, if the matrix-valued Boolean functions defined by them are the same: $F(f)=F(g)$. \end{defn}

\section{Symbolic Representation of Quantum States and Quantum Gates}

In this section, we show how matrix-valued Boolean expressions introduced in the above section can be used to describe quantum states and quantum gates. 

\subsection{Quantum States as Vector-Valued Boolean Expressions}

Let us first see how to represent a pure quantum state of qubits by a vector-valued Boolean expression. For each qubit variable $q$, we write $q^0=\overline{q}$ and $q^1=q$. Then an arbitrary state of $n$ qubits $q_1,...,q_n$:  
\begin{equation}\label{pure-state}|\psi\rangle=\sum_{i_1,...,i_n\in\{0,1\}}\alpha_{i_1\cdots i_n}|i_1,...,i_n\rangle\end{equation}
can be represented by a complex-valued Boolean expression:\begin{equation}\label{complex-state}\mu(|\psi\rangle)=\sum_{i_1,...,i_n\in\{0,1\}}\alpha_{i_1\cdots i_n}q_1^{i_1}\cdots q_n^{i_n}.\end{equation}
Note that $q_1,...,q_n$ are viewed as Boolean variables in (\ref{complex-state}). 
More generally, state (\ref{pure-state}) can also be represented as a vector-valued Boolean expression over a subset of $q_1,...,q_n$, say $q_{t_1},...,q_{t_k}$: 
\begin{equation}\label{vector-state}\mu(|\psi\rangle)=\sum_{i_{t_1},...,i_{t_k}\in\{0,1\}}\nu_{i_{t_1}\cdots i_{t_k}}q_{t_1}^{i_{t_1}}\cdots q_{t_k}^{i_{t_k}}
\end{equation}
where for each $(i_{t_1},...,i_{t_k})\in\{0,1\}^k$, $$\nu_{i_{t_1}\cdots i_{t_k}}=\left(\nu_{i_{t_1}\cdots i_{t_k}}(s)\right)_{s\in\{0,1\}^{n-k}}$$ is a $2^{n-k}$ dimensional column vector defined by
$$\nu_{i_{t_1}\cdots i_{t_k}}(s)=\alpha_{i_1\cdots i_n}$$ for each $s=(i_1,...,i_{t_1-1},i_{t_1+1},...,i_{t_k-1},i_{t_k+1},...,i_n)\in\{0,1\}^{n-k}.$

As we already saw in Example \ref{ex-in1}, expressions (\ref{complex-state}) and (\ref{vector-state}) can often be significantly simplified using Boolean reasoning. To further illustrate it, let us see two more examples:

{\vskip 3pt}

\begin{exam}\label{exam-uniform}The uniform superposition state of $n$ qubits $q_1,...,q_n$:\begin{equation}\label{uniform-1}|\psi\rangle=\frac{1}{\sqrt{2^n}}\sum_{i_1,...,i_n\in\{0,1\}}|i_1,...,i_n\rangle\end{equation} is represented by 
\begin{align}\mu(|\psi\rangle)& =\frac{1}{\sqrt{2^n}}\sum_{i_1,...,i_n\in\{0,1\}}q_1^{i_1}\cdots q_n^{i_n}\\
\label{uniform-2} &=\frac{1}{\sqrt{2^n}}(\overline{q}_1+q_1)\cdots (\overline{q}_n+q_n)\\ \label{uniform-3} &=\frac{1}{\sqrt{2^n}}.
\end{align}
It is worth noting that using Boolean equality $\overline{q}+q=1$ we are able to simplify (\ref{uniform-2}) to (\ref{uniform-3}), which is much more compact than (\ref{uniform-1}).

If we choose to use a Boolean expression over the first $k$ variables $q_1,...,q_k$, then $|\psi\rangle$ can be represented by 
\begin{align}\mu(|\psi\rangle)& =\frac{\nu}{\sqrt{2^k}}\sum_{i_1,...,i_n\in\{0,1\}}q_1^{i_1}\cdots q_k^{i_n}\\
\label{uniform-2} &=\frac{\nu}{\sqrt{2^k}}(\overline{q}_1+q_1)\cdots (\overline{q}_k+q_k)\\ \label{uniform-3} &=\frac{\nu}{\sqrt{2^k}}
\end{align} where $$\nu=\frac{1}{2^{n-k}}(1,...,1)^T$$ is a $2^{n-k}$ dimensional column vector representing the uniform superposition state of the last $n-k$ variables $q_{k+1},...,q_n$.  
\end{exam} 

{\vskip 3pt}
  
\begin{exam} The GHZ (Greenberger-Horne-Zeilinger) state of $n$ qubits $q_1,...,q_n$:
$$|{\rm GHZ}\rangle=\frac{1}{\sqrt{2}}(|0\rangle^{\otimes n}+|1\rangle^{\otimes n})$$ can be represented by the following Boolean expression:
\begin{align*}\mu(|{\rm GHZ}\rangle)&=\frac{1}{\sqrt{2}}(\overline{q}_1\cdots \overline{q}_n+q_1...q_n)\\ &=\frac{1}{\sqrt{2}}(q_1\leftrightarrow q_2)\cdots (q_{n-1}\leftrightarrow q_n).
\end{align*} 
\end{exam}

\subsection{Quantum Gates as Matrix-Valued Boolean Expressions} 

Now we turn to consider how to represent a quantum gate (i.e. a unitary transformation performed on qubits) by a matrix-valued Boolean expression. Let \begin{equation}\label{gate-1}U=\left(U_{ij}\right)_{2^n\times 2^n}\end{equation} be a unitary transformation on qubits $q_1,...,q_n$, where $i=i_1\cdots i_n, j=j_1\cdots j_n\in\{0,1\}^n$. We introduce a copy $q_i^\prime$ of $q_i$ for each $1\leq i\leq n$, and consider $q_1,...,q_n$ as the input variables and $q_1^\prime,...,q_n^\prime$ as the output variables. Then $U$ can be represented by the following complex-valued Boolean expression over $q_1,...,q_n,q_1^\prime,...,q_n^\prime$:
\begin{equation}\label{gate-2}\begin{split}\mu(U)=\sum_{i_1,...,i_n,j_1,...,j_n\in\{0,1\}}&U_{i_1\cdots i_n,j_1\cdots j_n}\\ &\ \cdot q_1^{i_1}\cdots q_n^{i_n}q_1^{\prime j_1}\cdots q_n^{\prime j_n}.\end{split}\end{equation}
In general, we can choose a subset $q_{t_1},...,q_{t_k}$ of $q_1,...,q_n$ and introduce their respective copies $q_{t_1}^\prime, ...,q_{t_k}^\prime$. Then quantum gate (\ref{gate-1}) can be represented by the following matrix-valued Boolean expression over $q_{t_1},...,q_{t_k},q_{t_1}^\prime, ...,q_{t_k}^\prime$:  
\begin{equation}\label{gate-3}\begin{split}\mu(U)=&\sum_{i_{t_1},...,i_{t_k},j_{t_1},...,j_{t_k}\in\{0,1\}} W_{i_{t_1}\cdots i_{t_k},j_{t_1}\cdots j_{t_k}}\\ &\qquad\qquad\qquad\cdot q_{t_1}^{i_{t_1}}\cdots q_{t_k}^{i_{t_k}}q^{\prime j_{t_1}}_{t_1}\cdots q_{t_k}^{\prime j_{t_k}}
\end{split}\end{equation}
where for each $(i_{t_1},...,i_{t_k})$ and $(j_{t_1},...,j_{t_k})\in\{0,1\}^k$, $$W_{i_{t_1}\cdots i_{t_k}, j_{t_1}\cdots j_{t_k}}=\left(W_{i_{t_1}\cdots i_{t_k},j_{t_1}\cdots j_{t_k}}(s,r)\right)_{s,r\in\{0,1\}^{n-k}}$$ is a $2^{n-k}\times 2^{n-k}$ matrix defined by
$$W_{i_{t_1}\cdots i_{t_k},j_{t_1}\cdots j_{t_k}}(s,r)=U_{i_1\cdots i_n,j_1\cdots j_n}$$ for each $s=(i_1,...,i_{t_1-1},i_{t_1+1},...,i_{t_k-1},i_{t_k+1},...,i_n)$ and $r=(j_1,...,j_{t_1-1},j_{t_1+1},...,j_{t_k-1},j_{t_k+1},...,j_n)\in\{0,1\}^{n-k}.$

An advantage of expressions (\ref{gate-2}) and (\ref{gate-3}) is also that they can often be significantly simplified by using Boolean logical reasoning, as shown in the following examples:   

{\vskip 3pt}

\begin{exam}The gate \begin{equation}\label{swap-1}{\rm SWAP}=\left(\begin{array}{cccc}1&0&0&0\\ 0&0&1&0\\ 0&1&0&0\\ 0&0&0&1\end{array}\right)\end{equation} 
is used to swap two qubits, say $q_1,q_2$. It can be described by the Boolean expression:
\begin{align}\mu({\rm SWAP})&=\overline{q}_1\overline{q}_2\overline{q}_1^\prime\overline{q}_2^\prime+\overline{q}_1q_2q_1^\prime\overline{q}_2^\prime+q_1\overline{q}_2\overline{q}_1^\prime q_2^\prime+q_1q_2q_1^\prime q_2^\prime\\
\label{swap-2}&=(q_1\leftrightarrow q_2^\prime)(q_2\leftrightarrow q_1^\prime).
\end{align} Obviously, expression (\ref{swap-2}) is closer to the intuition that $q_1$ and $q_2$ are swapped than matrix (\ref{swap-1}). 
\end{exam}

{\vskip 3pt}

\begin{exam}\label{Deutsch}The three-qubit Deutsch gate $D(\theta)$ performs the following transformation:
\begin{equation}\label{def-deutsch}|b_1,b_2,b_3\rangle \mapsto \begin{cases}&i\cos\theta |b_1,b_2,b_3\rangle+\sin\theta |b_1,b_2,1-b_3\rangle\\
 &\qquad\qquad\ \ \ {\rm if}\ b_1=b_2=1,\\
&|b_1,b_2,b_3\rangle\ \ \ {\rm otherwise}
\end{cases}
\end{equation} for any $b_1,b_2,b_3\in\{0,1\}$. 
In particular, when $\theta=\frac{\pi}{2}$, the Deutsch gate $D(\theta)$ becomes the Toffoli gate, which flips the third qubit, conditioned on the first two qubits being set on. 
The defining equation (\ref{def-deutsch}) of $D(\theta)$ can be straightforwardly translated into the complex-valued Boolean expression:
\begin{align*}\mu(D(\theta))=\ &[q_1q_2(i\cos\theta (q_3\leftrightarrow q_3^\prime)+\sin\theta (q_3\leftrightarrow \overline{q}_3^\prime))\\ &+(\overline{q}_1+\overline{q}_2)(q_3\leftrightarrow q_3^\prime)](q_1\leftrightarrow q_1^\prime)(q_2\leftrightarrow q_2^\prime)\\
=\ & [(i\cos\theta q_1q_2+\overline{q}_1+\overline{q}_2)(q_3\leftrightarrow q_3^\prime)\\ &+\sin\theta (q_3\leftrightarrow \overline{q}_3^\prime))](q_1\leftrightarrow q_1^\prime)(q_2\leftrightarrow q_2^\prime).
\end{align*}
\end{exam}

{\vskip 3pt}

\begin{exam} Let $U$ be a gate on $k$ qubits $p_1,...,p_k$. Then the controlled gate $C^n(U)$ is a unitary transformation on $n+k$ qubits $q_1,...,q_n,p_1,...,p_k$ defined by \begin{equation}\label{def-controlled}C^n(U)|b_1,...,b_n\rangle|\psi\rangle=|b_1,...,b_n\rangle U^{b_1\cdots b_n}|\psi\rangle 
\end{equation} for any $b_1,...,b_n\in\{0,1\}$ and $k$-qubit state $|\psi\rangle$. In particular, if $U$ is the rotation $$R_\theta=\left(\begin{array}{cc}i\cos\theta &\sin\theta\\ \sin\theta&i\cos\theta\end{array}\right)$$ of a single-qubit, then $C^2(U)$ is the Deutsch gate $D(\theta)$. Assume that $U$ is represented by complex or matrix-valued Boolean expression $\mu(U)$. Then $C^n(U)$ can be represented by the Boolean expression:
\begin{equation}\label{flexible-U}\begin{split}\mu\left(C^n(U)\right)&=\left[\left(\sum_{i=1}^n\overline{q}_i\right)\prod_{j=1}^k(p_j\leftrightarrow p_j^\prime)+ \prod_{i=1}^nq_i\cdot\mu(U)\right]\\ &\qquad\qquad\qquad\cdot\prod_{i=1}^n(q_i\leftrightarrow q_i^\prime).
\end{split}\end{equation} 
\end{exam}

It is worth pointing out that the choice of different subsets $q_{t_1},...,q_{t_k}$ of qubits in (\ref{vector-state}) and (\ref{gate-3}) gives us an opportunity to capture the regularity in quantum circuits and certain flexibility to identify appropriate granularity in representing and reasoning about them. This was witnessed in (\ref{uniform-3}) and (\ref{flexible-U}).    

\section{Symbolic Representation of Quantum Circuits}

In this section, we combine the ingredients introduced in the previous section to provide a symbolic representation of quantum circuits (with inputs). To simplify the notation, we slightly abuse the notation and use a matrix $U$, its matrix-valued Boolean expression $\mu(U)$ and the matrix-valued Boolean function $F(\mu(U))$ exchangeably.   

\subsection{Application of Quantum Gates}

We first consider the representation of the output state of a quantum gate. Let $|\psi\rangle$ be a pure state of qubits $q_1,...,q_n$ and $U$ a unitary transformation on $q_1,...,q_n$. Assume that $|\psi\rangle$ is given as a vector-valued Boolean expression over qubits $q_{i_1},...,q_{i_k}$ $(1\leq i_1<\cdots <i_k\leq n)$ and $U$ as a matrix-valued Boolean expression over $q_{i_1},...,q_{i_k},q_{i_1}^\prime,...,q_{i_k}^\prime$.
 
{\vskip 3pt}

\begin{thm}\label{thm-app} The application of $U$ on $|\psi\rangle$ yields: \begin{equation}\label{represent-app}U|\psi\rangle\equiv \exists q_{i_1},...,q_{i_k}:\mu(U)\cdot \mu(|\psi\rangle).
\end{equation} 
\end{thm} 

Here, $U|\psi\rangle$ should be understood as (the vector-valued Boolean expression) of a state of $q_1,...,q_{i_1-1},q_{i_1}^\prime,q_{i_1+1},...,q_{i_k-1},q_{i_k}^\prime,q_{i_k+1},...,q_n$. Of course, it can be seen as a state of $q_1,...,q_n$ modulo a substitution $[q_{i_1}/q_{i_1}^\prime,...,q_{i_k}/q_{i_k}^\prime].$

Examples \ref{ex-in1} and \ref{ex-in2} are simple instances of the above theorem. 

\subsection{Parallel Execution of Quantum Gates}

The parallel execution of two quantum gates $U$ and $V$ on two disjoint sets of qubits can be described as their tensor product $U\otimes V$. The following theorem shows that it can be represented by the product of the two matrix-valued Boolean expressions representing $U$ and $V$.  

{\vskip 3pt}

\begin{thm}\label{thm-tensor}Let $U$ and $V$ be unitary transformations on $q_1,...,q_n$ and $p_1,...,p_m$, respectively. If $\{q_1,...,q_n\}\cap\{p_1,...,p_m\}=\emptyset$, then \begin{equation}\label{represent-tensor} U\otimes V\equiv \mu(U)\cdot \mu(V).
\end{equation}
\end{thm}

{\vskip 3pt}

Note that if $\mu(U)$ is an expression over a subset $q_{t_1},...,q_{t_l}$ of $q_1,...,q_n$ and $\mu(V)$ over a subset $p_{s_1},...,p_{s_k}$ of $p_1,...,p_m$, then (\ref{represent-tensor}) is an expression over $q_{t_1},...,q_{t_l}, p_{s_1},...,p_{s_k}$ (and their output copies).   

To show how the above results can be used in reasoning about quantum circuits, let us see a multiple-qubit generalisation of Example \ref{ex-in1}.   

{\vskip 3pt}

\begin{exam} We consider a Hadamard transform $$H^{\otimes n}=\bigotimes_{i=1}^nH$$ on a register of $n$ qubits $q_1,...,q_n$. If $q_1,...,q_n$ are all initialised in basis state $|0\rangle$, by Theorems \ref{thm-app} and \ref{thm-tensor} we can compute the state of the register after the transformation: \begin{align*}  
H^{\otimes n}|0\rangle^{\otimes n}&\equiv\exists q_1,...,q_n: \frac{1}{\sqrt{2^n}}\prod_{i=1}^n\left[\left(\overline{q}_i+\overline{q}_i^\prime\right)-q_iq_i^\prime\right]\cdot \prod_{i=1}^n\overline{q}_i\\
&\equiv\exists q_1,...,q_n: \frac{1}{\sqrt{2^n}}\prod_{i=1}^n\left[\left(\left(\overline{q}_i+\overline{q}_i^\prime\right)-q_iq_i^\prime\right)\overline{q}_i\right]\\
&\equiv\exists q_1,...,q_n: \frac{1}{\sqrt{2^n}}\prod_{i=1}^n\overline{q}_i\\
&\equiv\frac{1}{\sqrt{2^n}}.
\end{align*}\end{exam} As shown in Example \ref{exam-uniform}, this is the uniform superposition (\ref{uniform-1}).
Here, the computation is accomplished by using Boolean logical laws $(q+q^\prime)q=q$, $q\overline{q}=0$ and $\exists q:\overline{q}=1$
rather than multiplication of large matrices. 

\subsection{Sequential Execution of Quantum Gates}

The sequential execution of two quantum gates $U$ and $V$ on the same set $q_1,...,q_n$ of qubits can be described by their product $UV$. For each $1\leq i\leq n$, we introduce two copies $q_i^\prime$ and $q_i^{\prime\prime}$ of $q_i$. Then we can assume that $U$ is represented by a matrix-valued Boolean expression over $q_1,...,q_n,q_1^\prime,...,q_n^\prime$, and $V$ by an expression over $q_1^\prime,...,q_n^\prime,q_1^{\prime\prime},...,q_n^{\prime\prime}$.
The following theorem gives a matrix-valued Boolean expression representation of $UV$ in terms of product and existential quantification. 

{\vskip 3pt}

\begin{thm}\label{thm-sequential}The sequential composition of $U$ and $V$: \begin{equation}\label{represent-sequential}UV\equiv \exists q_1^\prime,...,q_n^\prime: \mu(U)\cdot \mu(V).
\end{equation}
\end{thm}

{\vskip 3pt}

It is clear that (\ref{represent-sequential}) is an expression over $q_1,...,q_n,q_1^{\prime\prime},...,q_n^{\prime\prime}$. A simple application of the above theorem is presented in the following: 

{\vskip 3pt}

\begin{exam} Consider the square root of NOT gate on qubit $q$:
\begin{align*}\sqrt{{\rm NOT}}&=\frac{1}{2}\left(\begin{array}{cc} 1+i&1-i\\ 1-i&1+i\end{array}\right)\\ &=\frac{1+i}{2}(q\leftrightarrow q^\prime)+\frac{1-i}{2}(q\leftrightarrow\overline{q}^\prime).\end{align*}
Using Theorem \ref{thm-sequential}, we obtain:
\begin{align*}&\sqrt{{\rm NOT}}\cdot \sqrt{{\rm NOT}}\ =\ \frac{(1+i)^2}{4}\exists q^\prime:(q\leftrightarrow q^\prime)(q^\prime\leftrightarrow q^{\prime\prime})\\ 
&\qquad+\frac{1-i^2}{4}\exists q^\prime: [(q\leftrightarrow q^\prime)(q\leftrightarrow \overline{q}^{\prime\prime})+(q\leftrightarrow \overline{q}^\prime)(q\leftrightarrow q^{\prime\prime})]\\ 
&\qquad\qquad\qquad\qquad+\frac{(1-i)^2}{4}\exists q^\prime:(q\leftrightarrow \overline{q}^\prime)(q^\prime\leftrightarrow \overline{q}^{\prime\prime})\\ 
&=\frac{(1+i)^2+(1-i)^2}{4}(q\leftrightarrow q^{\prime\prime})+(q\leftrightarrow \overline{q}^{\prime\prime})\\ 
&=q\leftrightarrow \overline{q}^{\prime\prime}=\left(\begin{array}{cc} 0&1\\ 1&0\end{array}\right)={\rm NOT}. 
\end{align*} Here, we use the following Boolean logical laws: 
\begin{align*}&\exists q^\prime:(q\leftrightarrow q^\prime)(q^\prime\leftrightarrow q^{\prime\prime})=\exists q^\prime:(q\leftrightarrow \overline{q}^\prime)(q^\prime\leftrightarrow \overline{q}^{\prime\prime})=q\leftrightarrow q^{\prime\prime},\\
&\exists q^\prime:(q\leftrightarrow \overline{q}^\prime)(q^\prime\leftrightarrow q^{\prime\prime})=\exists q^\prime:(q\leftrightarrow q^\prime)(q^\prime\leftrightarrow \overline{q}^{\prime\prime})=q\leftrightarrow \overline{q}^{\prime\prime}. 
\end{align*}
\end{exam}

To summarise this section, a combination of Theorems \ref{thm-app}, \ref{thm-tensor} and \ref{thm-sequential} enables us to describe the behaviour of an arbitrary quantum circuit using a matrix-valued Boolean expression, which can be manipulated and simplified in the following way: \begin{itemize}\item Using Theorem \ref{thm-function}, the existential quantifications in (\ref{represent-app}) and (\ref{represent-sequential}) can be easily moved to the fronts of classical Boolean expressions; 
\item Then these existential quantifications over classical Boolean expressions can be computed by the techniques for more efficiently computing relational products in the literature of verification and symbolic model checking of classical circuits; for example, the technique of disjunctive and conjunctive partitioned transition relations described in Section V of \cite{Burch94}, which does not limit the size of the computed circuits.  
\end{itemize}

\section{Checking Equivalence of Quantum Circuits}

According to \cite{Molit}, from the industrial point of view, equivalence checking is the most important formal verification technique being employed in today's design flow of classical circuits. In this section, we demonstrate the applicability of the framework developed in the previous sections by using it in equivalence checking of quantum circuits. The main idea is to reduce the equivalence checking problem for quantum circuits represented as matrix-valued Boolean expressions to the same problem for classical Boolean expressions. Of course, operations of matrices  (including complex numbers and vectors) will be unavoidably involved in the process of reduction. But the size of these matrices should be significantly smaller than the dimension of the state space of the entire circuits under checking. 

The first step of our method for checking equivalence of two quantum circuits is to transform them into certain normal forms defined in the following:

{\vskip 3pt}  

\begin{defn}[Regular and Reduced Expressions] Let $f=\sum_{i=1}^lA_if_i$ be a matrix-valued Boolean expression. Then:  
\begin{enumerate}\item $f$ is called regular
if $A_i\neq 0$ for all $i=1,...,l$, and $f_1,...,f_l$ are pairwise contradictory; that is, $f_if_j=0$ for all $1\leq i<j\leq m$.
\item $f$ is called reduced if:\begin{enumerate}\item it is regular; and \item $A_i\neq A_j$ for all $1\leq i<j\leq m$. 
\end{enumerate}\end{enumerate}\end{defn}

{\vskip 3pt}

Obviously, the matrix-valued Boolean function $F=F(f)$ determined by a regular expression $f=\sum_{i=1}^mA_if_i$ can be easily computed as follows: 
$$F(b_1,...,b_n)=\begin{cases} A_i &{\rm if}\ f_i(b_1,...,b_n)=1,\\ 0 &{\rm otherwise.}
\end{cases}$$

We can use the following two rules for transforming matrix-valued Boolean expressions into reduced ones.   

\begin{itemize}\item \textbf{\textit{Regularisation Rule}}: There are many different ways of transforming an arbitrary matrix-valued Boolean expression to a regular one. For example, we can recursively use the following rule: if $\sum_{i=2}^nA_if_i$ is  regular, then
\begin{equation}\label{regularisation}\begin{split}&\sum_{i=1}^nA_if_i\Rightarrow A_1f_1\prod_{i=1}^n(\neg f_i)\\ &\qquad\qquad +\sum_{i=2}^n[A_i(\neg f_1)f_i+(A_1+A_i)f_1f_i].
\end{split}\end{equation}
\item \textbf{\textit{Reduction Rule}}: For any matrix $A$ and Boolean expressions $f_1,...,f_l$ that are pairwise contradictory: \begin{equation}\label{reduction}Af_1+\cdots +Af_l\Rightarrow A(f_1+\cdots +f_l).\end{equation}
\end{itemize}

It is easy to check that the above two rules are sound; that is, the two expressions before and after the transformation in (\ref{regularisation}) and (\ref{reduction}) are equivalent. Moreover, we observe that applying the reduction rule (\ref{reduction})  
to a regular expression always yields another regular expression. Thus, we have: 

{\vskip 3pt}

\begin{lem} Any matrix-valued Boolean expression can be transformed to a reduced expression by a finite number of applications of the regularisation and reduction rules. 
\end{lem}

{\vskip 3pt}

The following theorem then shows that equivalence checking of reduced matrix-valued Boolean expressions can actually be reduced to checking equivalence of classical Boolean expressions.  

{\vskip 3pt}

\begin{thm} Let $f=\sum_{i=1}^mA_if_i$ and $g=\sum_{j=1}^lB_jg_j$ be two reduced matrix-valued Boolean expressions over the same variables. Then $f\equiv g$ if and only if:\begin{enumerate}\item $m=l$; and \item there is a permutation $\sigma$ of $1,...,m$ such that 
$A_i=B_{\sigma(i)}$ and $f_i\equiv g_{\sigma(i)}$ for $i=1,....,m$.\end{enumerate}
\end{thm}

{\vskip 3pt}

Now we can check equivalence of quantum circuits expressed matrix-valued Boolean expressions $f$ and $g$ in the above theorem by employing various techniques and tools developed for equivalence checking of classical logic circuits to check whether $f_i\equiv g_{\sigma(i)}$ for $i=1,...,m$.   

\section{Conclusion}

This note describes a method for combining classical Boolean logical reasoning and matrix operations in reasoning about quantum circuits so that useful logical structure underlying the circuits can be captured and exploited.  
We only presented an application of this method in equivalence checking of quantum circuits in this note. But it seems that the method will be also useful in other tasks in the process of designing and implementing quantum computing hardware, including simplification, simulation, testing and verification of quantum circuits.     

In this note, we only considered combinational quantum circuits, but our approach can be easily extended for dealing with sequential quantum circuits \cite{Wang}. Indeed, we expect this approach can further help us in developing symbolic model checking techniques \cite{Burch92} for quantum systems \cite{YF21}.  

An implementation of this method will help us in verifying quantum circuits by employing the large variety of existing automatic tools for classical logic circuits (e.g. SAT solver, ROBDD). However, its effectiveness and efficiency are still unknown and need to be tested in future experiments on quantum circuits of a large size. We still do not understand in what circumstance our method can be really helpful, and also do not know how and to what extension it can be combined with other methods (e.g. those proposed in \cite{Rand, Hong, Pal18, Sei12, Shi, Via07, Markov}) for more efficient verification of quantum circuits.

\end{document}